\documentclass[aps,prl,twocolumn,superscriptaddress,showpacs]{revtex4}
\usepackage{amsmath,epsfig}
\usepackage{graphicx}
\newcommand{\ket}[1]{\mbox{$\left|#1\right\rangle$}}
\newcommand{\bra}[1]{\mbox{$\left\langle#1\right|$}}
\newcommand{\avg}[1]{\mbox{$\left\langle#1\right\rangle$}}
\newcommand{\op}[1]{\hat{#1}}
\newcommand{\kg}{\ket{g}}  %ground state
\newcommand{\bg}{\bra{g}}
\newcommand{\vdw}{\text{van der Waals}}
\newcommand{\vdwg}{V_g}
\newcommand{\vdwe}{V_e}
\newcommand{\dcg}{U_g}
\newcommand{\dce}{U_e}
\newcommand{\vect}[1]{{\bf{#1}}}
\newcommand{\sebare}{\Gamma}
\newcommand{\se}{\gamma}

\newcommand{\delt}{\delta}
\newcommand{\grad}{\vect{\nabla}}
\newcommand{\edc}{\vect{E}^\text{dc}}

\newcommand{\elas}{\vect{E}^\text{L}}
\newcommand{\elasm}{E^\text{L}_0}
\newcommand{\eps}{\epsilon}
\newcommand{\uac}{U_\text{L}}
\newcommand{\udc}{\dcg}%{U_\text{dc}}
\newcommand{\utrap}{U_\text{trap}}

\newcommand{\real}[1]{\text{Re}\left(#1\right)}

\newcommand{\bei}{\bra{e_i}}
\newcommand{\kex}{\ket{e_r}}
\newcommand{\key}{\ket{e_\phi}}
\newcommand{\kez}{\ket{e_z}}
\newcommand{\kel}{\ket{e}}
\newcommand{\bel}{\bra{e}}
\newcommand{\kep}{\ket{e_\bot}}

\newcommand{\isum}{i=r,\phi,z}
\newcommand{\isumrz}{i=r,z}
\newcommand{\ijsumrz}{i,j=r,z}
\newcommand{\ijksumrz}{i,j,k=r,z}
\newcommand{\ketr}{\ket{e_r}}

\newcommand{\epsz}{\epsilon_z}
\newcommand{\epsr}{\eps_r}

\begin{document}
\title{Electro-Optical Nanotraps for Neutral Atoms}

\author{Brian Murphy}
\affiliation{School of Engineering and Applied Sciences, Harvard
University, Cambridge, Massachusetts 02138}

\author{Lene Vestergaard Hau}
\affiliation{School of Engineering and Applied Sciences, Harvard
University, Cambridge, Massachusetts 02138} \affiliation{Department
of Physics, Harvard University, Cambridge, Massachusetts 02138}

\begin{abstract}
We propose a new class of nanoscale electro-optical traps for
neutral atoms.  A prototype is the toroidal trap created by a
suspended, charged carbon nanotube decorated with a silver
nanosphere dimer. An illuminating laser field, blue detuned from an
atomic resonance frequency, is strongly focused by plasmons induced
in the dimer and generates both a repulsive potential barrier near
the nanostructure surface and a large viscous damping force that
facilitates trap loading. Atoms with velocities of several meters
per second may be loaded directly into the trap via spontaneous
emission of just two photons.
\end{abstract}

\pacs{37.10.De, 37.10.Gh, 73.22.Lp, 78.67.Bf} \maketitle

Laser cooling and trapping of neutral atoms have led to remarkable
successes including the creation of Bose-Einstein condensates
\cite{CornellNobelLectureRMP2002}, and a quest to miniaturize and
chip-integrate trapping and cooling structures has represented an
extremely active research area in recent years
\cite{ZimmermannMagneticMicroRMP2007}. Cold atoms have successfully
been trapped in micron-sized traps but efforts to push the
technology to the nanoscale have been unsuccessful. Proposals for
atomic traps near nanostructures have been put forward
\cite{KlimovDots2003}; however, methods to overcome the attractive
Van der Waals forces near material surfaces
\cite{KnightNanotube2007} as well as the heating caused by thermally
induced charge and current fluctuations in room temperature
materials in close proximity to trapped atoms
\cite{HenkelPotting1999Jones2003Cornell2003} have not yet been
proposed. Here we present an electro-optical nanotrap that overcomes
these obstacles and, importantly, also provides for damping of
atomic motion that facilitates direct trap loading.

The strong coupling of a trapped atom to the nanoscopic structure
allows for sensitive probing of electromagnetic fields and
dielectric properties of materials at the nanoscale, of great
importance for the flourishing field of single biomolecule studies
with surface enhanced Raman scattering \cite{BrusSERSrev2003}. It
also facilitates studies of QED effects in atom-surface
interactions, a topic of intense current interest for applications
to nanomechanical devices and in the search for forces beyond the
standard model \cite{AntezzaStringari2004}.

Metallic nanoparticles exhibit plasmon resonances. An illuminating
electromagnetic field can excite plasma oscillations that generate
large and localized electric fields near the particles
\cite{BrusSERSrev2003}.  Here, we consider the effect of such fields
on atomic motion. A plasmon-resonant laser field that is
blue-detuned from an atomic resonance causes a dramatic repulsion of
atoms from the nanostructure surface, and when combined with
attractive electrostatic forces creates a trap minimum at nanometer
distances from that surface. Simultaneously, large gradients of the
plasmon-enhanced fields provide a strong dissipative force.

%Idea for trap

%figure 1 here
\begin{figure}[]
\begin{center}
\includegraphics[width=8.6cm]{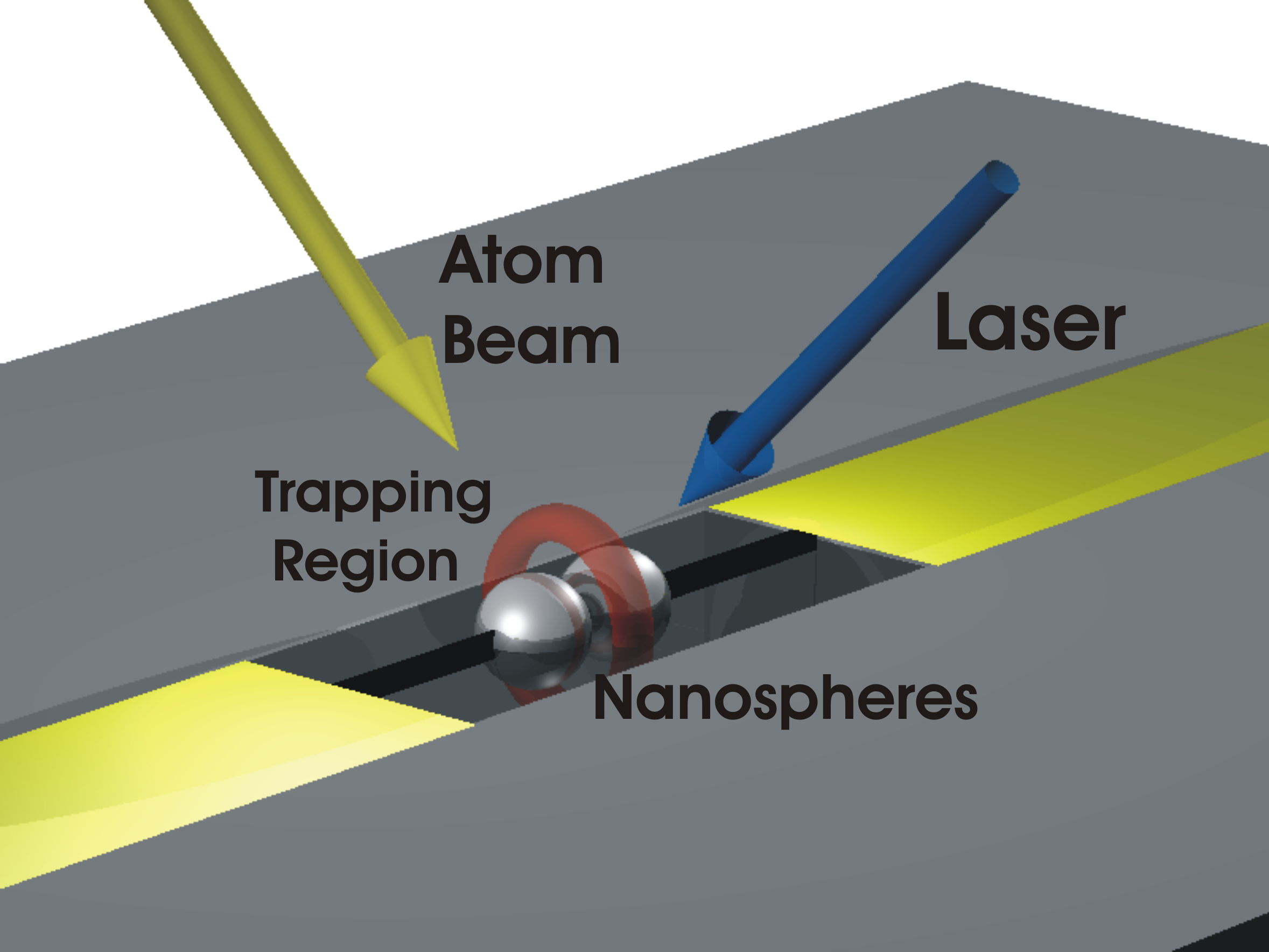}
\end{center}
\caption{(Color) The nanotrap.  A suspended carbon nanotube,
attached at each end to an electrode, supports two Ag nanospheres
across a gap in a silicon nitride membrane. An illuminating laser
field excites plasma oscillations in the spheres and large electric
fields near the structures are generated. In addition, a dc voltage
is applied to the electrodes to create a toroidal trapping region
(red). The radius of the toroid can be controlled with nanometer
precision and the trap may be loaded directly from an incident atom
beam.} \label{chipdrawingfigure}
\end{figure}

As diagrammed in Fig. \ref{chipdrawingfigure}, a nanotrap is formed
by two Ag spheres that are supported and charged by a carbon
nanotube. Suspended nanotubes with lengths of many microns have been
fabricated across gaps \cite{PengGaps2003}.

As described in detail below, the trapping potential for an atom
near the nanospheres is approximated by
\begin{eqnarray}
\utrap & = & \uac + \udc + \vdwg \label{approximateU}
\end{eqnarray}
where $\udc$ is the dc Stark shift and $\vdwg$ the $\vdw$ energy of
the ground state. Also, $\uac = \frac{\hbar\delt}{2}\ln(1+s)$ is the
optical dipole potential \cite{Ashkin1978,DalibardDressed1985} due
to the plasmon-enhanced laser field, where
$s=\frac{\Omega^2/2}{\se^2/4+\delt^2}$ is the saturation parameter,
$\delt$ is the detuning between the laser and the atomic resonance
including dc Stark shifts and $\vdw$ shifts, $\se$ is the decay rate
of the excited atomic state modified by the presence of the
nanostructure as discussed later in the text, and $\Omega =
\frac{\mu \elasm}{\hbar}$ is the Rabi frequency representing the
coupling of the plasmon enhanced electric field of amplitude
$\elasm$ to the atomic dipole moment $\mu$.

%figure 2 here

\begin{figure}[]
\begin{center}
\includegraphics[width=8.6cm]{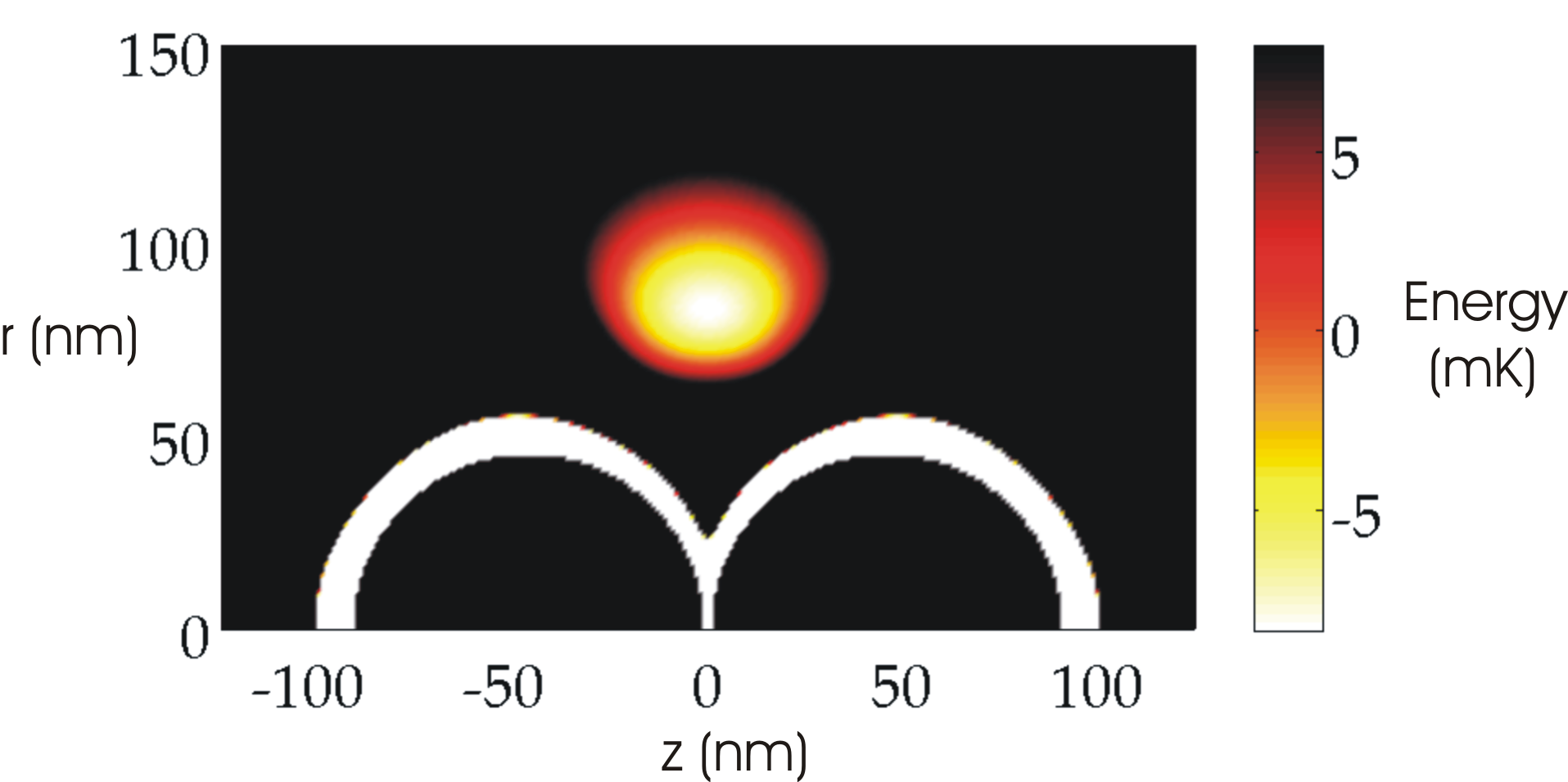}
\end{center}
\caption{(Color) The trap potential (\ref{approximateU}). The $z$
axis is parallel to the nanotube, and $r$ is the distance from the
tube. Incident laser light of wavelength $\lambda = 475$nm is
polarized along the z-axis connecting two Ag spheres of diameter 90
nm and separation 2nm. The plasmon resonance of this dimer is at
457nm. The detuning between the laser frequency $\omega$ and the
bare atomic resonance $\omega_0$ is picked to be
$\Delta=\omega-\omega_0=500\Gamma$, where
$\Gamma=\frac{q^2\omega_0^2}{6\pi\varepsilon_0 m_e c^3}$ is the
decay rate of the excited atomic state in free space.  The incident
laser intensity is $I^\text{inc}=4.6\times 10^3$W$/$cm$^2$, and the
nanospheres are charged to a voltage of $V=3.5$ Volts (relative to a
macroscopic grounding surface).  With these parameters, the
saturation parameter $s=0.09$ at the trap minimum.  For contrast,
the color scale has a lower limit corresponding to the minimum value
of $\utrap$ in the trapping region and a range equal to the
equilibrium temperature. The low energy area near the surface of the
nanospheres is the region where the attractive $\vdw$ force
dominates, and is not part of the trapping region. The trapping
potential has cylindrical symmetry in the quasi-static limit.  When
the voltage is turned off, a trapping potential persists due to a
minimum in the laser field amplitude at $r=94$nm, but the barrier
for escape is reduced in this case from $48$ to $29$mK.  }
\label{trappotentialfigure}
\end{figure}

To solve for both the static and laser-induced electric fields in
the near field, we work in the quasi-static limit. For an incident
plane wave laser field $\vect{E}^\text{inc}$ at frequency $\omega$,
the resulting electric field is $E^\text{L}_j(\vect{r}) =
M_{ji}(\vect{r},\omega)E^\text{inc}_i$, where $M$ describes the
local, plasmon-induced enhancement. For an external dipole source
$\vect{p}$ oscillating at frequency $\omega$ at location
$\vect{r}'$, the induced charge distribution in the nanostructure
creates a reflected electric field ${E}^\text{ref}_j(\vect{r}) =
G_{ji}(\vect{r},\vect{r}';\omega)p_i$, where $G$ is the tensor
Green's function for the given geometry. The dc electric field
$\edc$ is calculated with the spheres fixed at a particular voltage,
and in all three cases we use the dielectric constant
$\varepsilon_\text{Ag}$ of Ag at the relevant source frequency
\cite{PalikV2}.  Figure \ref{trappotentialfigure} shows the
resulting trapping potential (\ref{approximateU}) for an atom with
unit oscillator strength and a resonance frequency close to the
plasmon resonance of the dimer-tube structure.

The illuminating laser will cause heating in the nanospheres
at a rate of $2\mu$W per sphere, resulting in a dimer temperature of $\approx 640$K which is well below
the melting point of silver. Here we assume a 10 $\mu$m long, 4nm diameter
nanotube fixed to a $300K$ membrane \cite{DaiThermalModel2005}.

We model an atom's internal structure with a single ground state
$\kg$ and three excited states $\kex$,$\key$, and $\kez$ that are
degenerate for a free space atom. The dipole matrix element
is $\bei \op{\vect{d}} \kg = \mu \hat{i}$ where
$\{\hat{i}\}_{\isum}$ form orthonormal basis vectors in a
cylindrical coordinate system, $\mu=\sqrt{\frac{\hbar
q^2}{2m_e\omega_0}}$ (the value for a two-level atom), $q$ is the
electron charge, $m_e$ is the mass of the electron,
and $\omega_0$ is the resonance frequency of the atom.  The plasmon
enhanced electric field of the laser is $\elas =
\frac{1}{2}\elasm\hat{\eps}e^{-i\omega t}+c.c.$, where $\elasm$ and
$\hat{\eps}$ are the position-dependent amplitude and polarization
vector.  We change two of the excited state basis vectors to
%\begin{equation}
$\kel  = \epsr\ketr + \epsz\kez$ and $\kep  =  \epsz^*\ketr - \epsr^*
\kez$.
%\end{equation}
The Hamiltonian (in the rotating frame) governing the internal
dynamics of an atom interacting with the charged and
laser-illuminated nanotube-dimer system is then
\begin{eqnarray}
\op{H} & = &  \left(\vdwg+\dcg\right)\kg\bg +
\left(-\hbar\Delta+\vdwe+\dce\right)\kel\bel
\nonumber \\
&& -\frac{\hbar\Omega}{2}\left(\kel\bg + \kg\bel\right).
\label{hamiltoniantwolevel}
\end{eqnarray}
Here $\vdwe$ and $\dce$ are the van der Waals and dc Stark shifts of
the excited state $\kel$, and we have used the rotating wave
approximation and the fact that the laser couples only states $\kg$
and $\kel$ to an excellent approximation. From
(\ref{hamiltoniantwolevel}), we find the force on an atom at rest to
be
\begin{eqnarray} \avg{\vect{F}} & = &
-\frac{\hbar\delt\Omega\grad\Omega}{2\left(1+s\right)\left(\se^2/4+\delt^2\right)} \nonumber \\
&& + \frac{\hbar
s\grad\delt}{2\left(1+s\right)}-\grad\left(\vdwg+\dcg\right)
\label{force0}
\end{eqnarray}
where $\delta = \Delta - \left(\vdwe+\dce-\vdwg-\dcg\right)/\hbar$
and $\Delta=\omega-\omega_0$. In the regime of large and relatively
constant detuning ($\delt>>\se$ and $ |\frac{\nabla\delt}{\delt}| <<
\frac{|\nabla\Omega|}{\Omega}$ ), the force in (\ref{force0}) is
minus the gradient of the trap potential in (\ref{approximateU})
which is plotted in Fig. \ref{trappotentialfigure}.

The dc Stark shifts of the ground and excited states are
%\begin{equation}
$\dcg =  -\frac{1}{2}\alpha_0|\edc|^2$ and $\dce  =
\frac{1}{2}\alpha_0|\edc\cdot\hat{\eps}|^2$,
%\end{eqnarray}
where the static polarizability is given by $\alpha_0 =
\frac{q^2}{m_e\omega_0^2}$. (Since $\Delta
>> \dce/\hbar$ and $s\ll 1$ in the trap region for the parameters used here, the atom dynamics do not depend on the
specific form of $\dce$.)

For a nanoscale trap, $\vdw$ (and Casimir-Polder
\cite{CasimirPolder1948}) shifts and modifications to the excited
state lifetime are important. In the near field, the ground state
energy shift of an atom in front of a perfect, planar conductor is
the London energy of a fluctuating dipole moment interacting with
its image dipole. Here we approximate the $\vdw$ shift of
the ground state by the energy of a fluctuating dipole in front of
two conducting spheres,
%\begin{equation}
$\vdwg \approx -\frac{1}{2}|\mu|^2\sum_{\isum} G_{ii}^\text{stat}$,
%\label{vdwg}
%\end{equation}
where $G_{ii}^\text{stat}=G_{ii}(\vect{r},\vect{r};\omega=0)$.

Excited state $\vdw$ shifts have two contributions, of which one is
an inverted ground state $\vdw$ shift and the other is due
to a modification of the resonance frequency of a Lorentz oscillator
from the oscillator's self-interaction with the induced, reflected
field from the nanostructure \cite{CPS1978}. Hence
%\begin{equation}
$\vdwe \approx
\frac{1}{2}|\mu|^2\sum_{\isumrz}|\eps_i|^2\left(G_{ii}^\text{stat}-2\real{G_{ii}^\text{res}}\right)$,
%\end{equation}
where $G_{ii}^\text{res}=G_{ii}(\vect{r},\vect{r};\omega_0)$ \cite{HindsSimple1991}.

The decay rate of the excited state is also modified
\cite{Purcell1946} due to the proximity of the atom to the
nanostructure. The decay rate calculated from the damping of a
classical, oscillating dipole interacting with its induced electric
field from the nanostructure agrees with a full quantum treatment
\cite{CPS1978}. In the quasi-static limit, this damping yields only the nonradiative
decay rate corresponding to energy loss in the nanostructure
\cite{KlimovNanobodies2003} according to
\begin{equation}
\frac{\se_{nr}}{\sebare} =
\frac{6\pi\varepsilon_0}{k^3}\sum_{\ijsumrz}\text{Im}\left(\eps_i\eps_j^*
G_{ji}^\text{res}\right), \label{gammanonrad}
\end{equation}
where $\Gamma$ is the free-space decay rate. The radiative decay $\se_r$ can then be added separately.  From
energy conservation, we find $\gamma_r$ by calculating the energy
radiated into the far field.  We can relate the dipole far field
emission to the local field enhancement of an incident plane wave by
appealing to the reciprocity theorem \cite{EtchegoinORT2006} and
find:
\begin{equation}
\frac{\se_r}{\sebare} = \sum_{\ijksumrz}\eps_i\eps_j^*M_{ik}M_{jk}^*
\label{gammarad}
\end{equation}
We then obtain $\se=\se_r+\se_{nr}$ everywhere in space from
(\ref{gammanonrad}) and (\ref{gammarad}).

%Trap design w/ cooling
For an atom to be stably trapped, the equilibrium temperature
characterizing the average kinetic energy of its center-of-mass
motion must be less than the potential barrier for trap escape.  A
strong blue-detuned laser field leads not only to trapping
(\ref{force0}) but also to damping \cite{DalibardDressed1985}.  In
the "strong-field limit" where $\Omega$,$\delt$,$s\Omega$,$s\delt >>
\se$, and to first order in velocity, the damping of the
center-of-mass motion is given by the force
\begin{equation}
\vect{F} \approx -\frac{\hbar s^2}{\delt\gamma(1+s)^3}\left((\nabla
\Omega)\cdot\vect{v}\right)\nabla \Omega\label{betaeq}
\end{equation}
Its dependence on the gradient of the Rabi frequency leads to
extraordinarily large damping rates with the strongly localized
fields of the nanotube-dimer trap. (The linear velocity dependence
of the damping force is valid for velocities below a critical
velocity $v_c \approx  \se\Omega/\left|\grad\Omega\right|$).
Furthermore, an atom is heated due to fluctuations of the trapping
force from spontaneous transitions associated with the damping. The
heating is characterized by a momentum diffusion coefficient $D$,
and in the strong-field limit, $D \approx \frac{\hbar^2
s^3}{2\gamma(1+s)^3}|\nabla \Omega|^2$ \cite{DalibardDressed1985}.
The equilibrium temperature for atomic motion is given by the ratio
between $D$ and the damping (velocity) coefficient of
(\ref{betaeq}), averaged over an isotropic velocity distribution, so
$k_BT = \frac{\hbar\Omega^2}{4\delta}$.  With the parameters of Fig.
\ref{trappotentialfigure}, this would lead to an equilibrium
temperature of 16mK which is below the potential barrier of 48mK for
trap escape.

%figure 3

\begin{figure}[]
\begin{center}
\includegraphics[width=8.6cm]{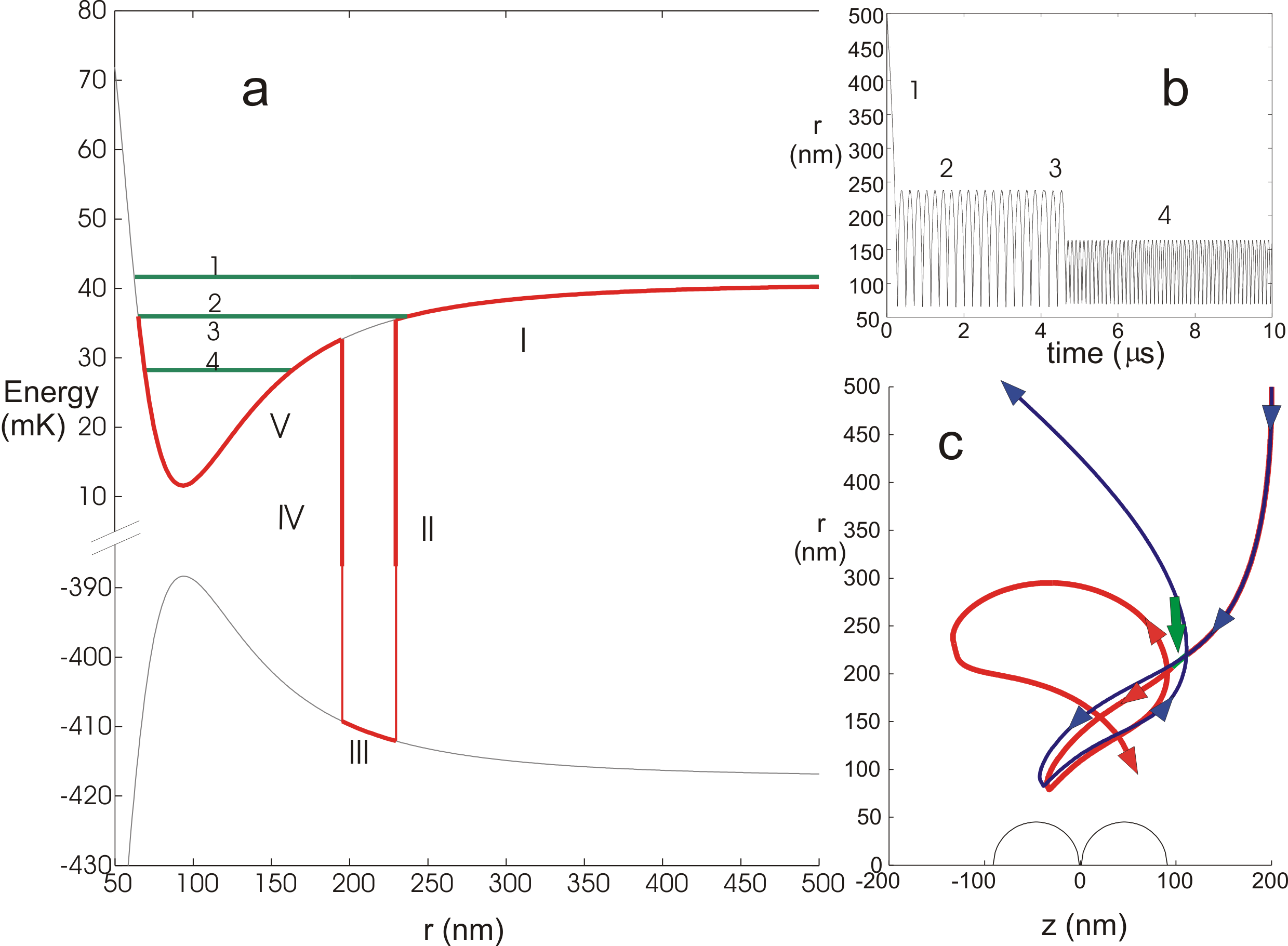}
\end{center}
\caption{(Color) Atom trajectories studied with Monte Carlo
simulations. In (a) we show a typical capture process for an atom
launched along the $z=0$ symmetry axis with velocity 1 m/s at
$r=500$nm (corresponding to a launch velocity of 5.5 m/s from
infinity). The dressed state energies are plotted as gray curves
(top curve is for $\ket{1}$  and bottom curve for $\ket{2}$), and
the state actually occupied by the atom is marked in red (and
labelled I, III, V) (In the limit of vanishing laser field $\ket{1}$
($\ket{2}$) is simply $\ket{g}$ ($\ket{e}$) but for non-zero field,
the dressed states are superpositions of $\ket{g}$ and $\ket{e}$
\cite{DalibardDressed1985}).  In regions I and V the atom is in
dressed state $\ket{1}$, in region III it is in state $\ket{2}$, and
after transitions II and IV the particle has suffered a net energy
loss. The energy dissipation in such processes is provided by loss
in the nanoparticle and by emission of fluorescence photons with
frequencies up or down shifted relative to the laser frequency by
amounts given by the energy differences between the dressed states.
In (b) the same atomic motion is followed further, for up to
10$\mu$s. The successive stages, corresponding to different energies
reached after each pair of transitions, are labeled 1-4 in both (a)
and (b). In (c), Monte Carlo trajectories are plotted for both
uncaptured (blue) and captured (red) atoms launched with velocity 1
m/s from (r,z)=(500,200)nm (The paths are followed for 734 ns and
1.1$\mu$s, respectively). The captured atom undergoes a pair of
spontaneous transitions in the region indicated by a separate arrow
(green) \label{loadingfigure}}
\end{figure}

Atoms are trapped with a transverse oscillation period of 50ns. This
is on the order of the inverse spontaneous decay rate $\gamma$,
which leads us to re-examine the microscopic nature of the damping
mechanism. We perform Monte Carlo simulations of atom trajectories,
and for motion along the $z=0$ symmetry axis we show in Fig.
\ref{loadingfigure} how an atom spontaneously transitions between
the energy curves corresponding to the two ``dressed'' eigenstates,
$\ket{1}$ and $\ket{2}$, of the Hamiltonian
(\ref{hamiltoniantwolevel}) \cite{DalibardDressed1985}. (In our
case, the eigenstates are ``dressed" by plasmons and are determined
by the coupling between the atom and the longitudinal, plasmon
induced electric field). We find that the time $\tau$ an atom
remains trapped in dressed state $\ket{1}$ increases when the
voltage is turned off and can be tuned over many orders of
magnitude.  We find $\tau$=25$\mu$s for $\Delta=500\Gamma$, whereas
$\Delta=6000\Gamma$ results in $\tau=900$ms. Heating due to Johnson
noise in the nanospheres is negligible on this time scale. This is
very different than for magnetic microtraps where large amounts of
conducting material are in close proximity to trapped atoms
\cite{HenkelPotting1999Jones2003Cornell2003}.

Monte Carlo simulations have also been used to study the loading of
atoms into the nanotrap.  We find that an atom launched towards the
nanostructure with a velocity of meters per second may lose enough
energy to become trapped after a single pair of spontaneous
transitions between the dressed states. Figure \ref{loadingfigure}a
shows how an incoming atom in state $\ket{1}$ transitions to
$\ket{2}$ and then loses kinetic energy as it climbs ``uphill'' in
this state. After the reciprocal transition
$\ket{2}\rightarrow\ket{1}$, the atom remains trapped in $\ket{1}$,
having lost potential and kinetic energy in equal amounts.  The
energy is dissipated via the atom's coupling to plasmon oscillations
that are subsequently damped both by coupling to the transverse
(radiating) electromagnetic field modes and by loss in the
nanostructure. In Fig. \ref{loadingfigure}b we continue to follow
the motion of the trapped atom for 10 $\mu$s. These simulations do
not rely on Eqs (\ref{force0}) and (\ref{betaeq}).

The nanotrap can be loaded by atoms launched at a few meters per
second from a high-density magneto-optical trap. We estimate that
with nanotrap parameters as in Fig. 2 (except with a nanotube
voltage of $V=0$), a magneto-optical trap density of
$10^{12}$cm$^{-3}$, a launch velocity of 5.5 m/s, and a longitudinal
temperature of 200$\mu$K \cite{Cornell1994}, there is an
approximately 2\% probability for atom capture within a
cross-sectional area of 450nm ($\Delta$z) by 800nm (Fig.
\ref{loadingfigure}c). This results in an average loading time of
25$\mu$s, and after a brief loading stage, the laser detuning may be
increased to maximize the trap lifetime. To match the plasmon
resonance with a particular atomic resonance, the nanotube may be
decorated with dimers of nanoshells rather than solid silver spheres
\cite{HalasNanoengineering1998}. Surface adsorption of atoms is
prevented with heated silver spheres as used here (and for sodium
atoms, by an ionization potential that is much larger than the
workfunction for silver) \cite{Cornell2007Adsorbates}.

We have described a novel nanotrap for neutral atoms. The trap
frequency is several MHz, the ground state is transversely localized
to within a few  nm, and trap lifetimes exceed $10^6$ oscillation
periods. Atoms can be held in multiple hyperfine states facilitating
sideband cooling to the ground state \cite{MonroeSideband1995}.
Spectroscopic determination of the trap's energy levels, with
parts-per-million precision, is possible with rf fields applied
directly to the nanotube. The trap minimum is in the crossover
region between the Casimir-Polder and the Van der Waals-London
regimes \cite{AntezzaStringari2004}, and in the limit of large laser
detuning, nanotrap studies could cast new light on the interplay
between quantum fluctuations in the radiation field, the atom, and
the nanostructure \cite{DalibardCohenT1982}, with a proper
description ultimately involving entangled eigenstates of the
strongly coupled atom-nanostructure system. Atom-surface
interactions have recently been a subject of intense studies that
have opened new possibilities for testing the existence of
non-Newtonian gravitational forces at small length scales
\cite{Geraci2003Cornell2005}. Studies based on the nanotrap have the
potential to push such tests to length scales of 100 nm and below, a
regime inaccessible in prior experiments.

This work was supported by the Air Force Office of Sponsored
Research.

%%%%%%%%%%%%%%%%%%%%%%%%%%%%%%%%%
%\bibliography{hybridtrap071008,transfer-paper-v5}

\end{document}